\documentclass[twocolumn,prl,superscriptaddress,showpacs,preprintnumbers,amsmath,amssymb]{revtex4}

\usepackage{graphicx}
\usepackage{subfigure}
\usepackage{dcolumn}
\usepackage{bm}

\begin{document}

\title{Plasma surface dynamics and smoothing in the relativistic few-cycle regime}

\author{S. G. Rykovanov}
    \email[Corresponding author: ]{sergey.rykovanov@mpq.mpg.de}
    \affiliation{Max-Planck-Institut f\"ur Quantenoptik,  D-85748 Garching, Germany}
    \affiliation{Moscow Engineering Physics Institute, 115409 Moscow, Russia}
    \affiliation{Fakult\"at f\"ur Physik, Ludwig-Maximilians-Universit\"at M\"unchen, D-85748 Garching, Germany}
\author{H. Ruhl}
    \affiliation{Fakult\"at f\"ur Physik, Ludwig-Maximilians-Universit\"at M\"unchen, D-85748 Garching, Germany}
\author{J. Meyer-ter-Vehn}
    \affiliation{Max-Planck-Institut f\"ur Quantenoptik,  D-85748 Garching, Germany}
\author{R. H\"{o}rlein}
    \affiliation{Max-Planck-Institut f\"ur Quantenoptik, D-85748 Garching, Germany}
    \affiliation{Fakult\"at f\"ur Physik, Ludwig-Maximilians-Universit\"at M\"unchen, D-85748 Garching, Germany}
\author{B. Dromey}
    \affiliation{Department of Physics and Astronomy, Queens University Belfast, BT7 1NN, UK}
\author{M. Zepf}
    \affiliation{Department of Physics and Astronomy, Queens University Belfast, BT7 1NN, UK}
\author{G. D. Tsakiris}
    \affiliation{Max-Planck-Institut f\"ur Quantenoptik,  D-85748 Garching, Germany}

\date{\today}

\begin{abstract}

In laser-plasma interactions it is widely accepted that a non-uniform interaction
surface will invariably seed hydrodynamic instabilities and a growth in the
amplitude of the initial modulation. Recent experimental results [Dromey, Nat. Phys.
2009] have demonstrated that there must be target smoothing in femtosecond timescale
relativistic interactions, contrary to prevailing expectation. In this paper we develop a
theoretical description of the physical process that underlies this novel phenomena.
We show that the surface dynamics in the few-cycle relativistic regime is dominated
by the coherent electron motion resulting in a smoothing of the electron surface. This
stabilization of plasma surfaces is unique in laser-plasma interactions and
demonstrates that dynamics in the few-cycle regime differ fundamentally from the
longer pulse regimes.  This has important consequences for applications such as
radiation pressure acceleration of protons and ions and harmonic generation from
relativistically oscillating surfaces.

\end{abstract}

\maketitle

The dynamics of the plasma-vacuum surface are of critical importance to the understanding laser-plasma interactions.
Until now, the prevailing expectation has been that laser-plasma surfaces are inherently unstable entities, since the morphology of the laser-plasma interaction surface is determined by hydrodynamics for timescales where ion motion is significant and initial perturbation will grow rapidly due to instabilities.  Specifically, the surface dynamics play a central role for many applications of intense laser-plasma interactions. The best known example is probably the Rayleigh-Taylor instability which limits the parameter space that can practically be accessed in Inertial Confinement Fusion \cite{Lindl95}. For relativistic pulses with durations of $>$100fs hydrodynamic motion is also the key
ingredient to determining the surface morphology. In this regime the role of the light-fluid is performed by the
ponderomotive pressure of an intense laser \cite{WilksPRL92, ZepfPOP96}, in close analogy to the Rayleigh-Taylor instability. The growth and evolution of this instability is essential to understanding hole-boring \cite{TabakPOP94, WillingalePRL09}. The sole exception to this view of the interaction-surface dynamics were ultrashort, few-cycle interactions at modest intensities where there is insufficient time for ion motion to take place and consequently the initial surface shape is typically assumed to remain unchanged during the interaction.

However, recent results by Dromey \emph{et al.} showed that the reflection of high-order harmonics from initially rough targets was consistent with the existence of a very effective smoothing mechanism \cite{Dro08Div, Dro07KeV}. This suggests a new paradigm in laser plasma interactions, in that the exact opposite of the generally accepted behaviour occurs $-$ surface smoothing rather than a modulation growth due to instabilities.
Motivated by this surprising behaviour, we have investigated the interaction of a relativistically strong laser pulse with an overdense, modulated plasma surface. We show for the first time that the surface morphology in the few-cycle relativistic regime is dominated by coherent electron motion in the laser field. This is a complete departure from physical picture of how the interaction surface evolves in laser-plasma interactions and has important consequences for the frontier of ultra-fast science in the relativistic regime. This effect exists in a broad, practically important parameter range spanning ultrafast pulses with intensities from 10$^{18}$Wcm$^{-2}$ upwards.

To gain insight  into the particle motion in the over-dense plasma we use a simple one-dimensional model \cite{Wilks93, Zaretsky04, Mulser08}. This single-particle model describes the motion of an incompressible electron layer bound to  immobile ion background via charge-separation fields under the influence of normally incident, linearly polarized electromagnetic wave. This layer, later referred to simply as the electron, is initially located on the vacuum-plasma interface at $x=0$. Taking into account that the charge separation fields are proportional to the electron longitudinal coordinate $x $, the equations of motion can be readily obtained as:

\begin{align}
\frac{dp_x}{dt}&=-\beta_y \frac{\partial a_y(t,x)}{\partial x}+n_ex,
\label{subeqmotion-a}\\
\frac{dp_y}{dt}&=\frac{da_y(t,x)}{dt}.
\label{subeqmotion-b}
\end{align}

where $x,y$ are the propagation and transversal coordinates respectively, $\beta_{x,y}$ and $p_{x,y}$ - velocity and momenta components respectively, $a_y$ - the driving vector potential and $n_e$ is the electron density. We work in \emph{relativistic units}. The normalized quantities for vector potential $a$, time $t$, length $l$, momentum $p$, and density $n$ are obtained
from their counterparts in SI-units $A$, $t'$, $l'$, $p'$, and
$n'$ via
\begin{equation}
a = \frac{eA}{m_ec},~t=\omega_Lt', ~l=\frac{\omega_L}{c}l',
~p=\frac{p'}{m_ec}, ~n = \frac{n'}{n_{cr}} \mathrm{.}
\end{equation}
Here $e$ and $m_e$ are the charge and the mass of the electron,
$\omega_L$ is the laser angular frequency, $c$ is the speed of
light in vacuum, and $n_{cr}= \varepsilon_0m_e\omega_L^2/e^2$
is the electron critical density.



In equations (\ref{subeqmotion-b}) $a_y(t,x )$ denotes the driving vector potential on the vacuum-plasma interface, which results from the interference between the incident and reflected wave. It can be found by imposing the standard boundary conditions for the continuity of electromagnetic fields, i.e. of vector potential and its spatial derivative $\partial_x a_y$ at the plasma-vacuum interface. Without losing the generality, the incident $a^i$, reflected $a^r$, and transmitted $a^t$ vector potentials can be taken in the form $a_y^i(t-x^\prime)=-E_i\cdot\sin(t-x^\prime)$, $a_y^r(t+x^\prime)=-E_r\cdot\sin(t+x^\prime+\phi_r)$ and $a_y^t(t,x^\prime)=-E_t\cdot\sin(t+\phi_t)\cdot\exp[-\omega_p(x^\prime-x )]$ respectively. With $x^\prime$ we denote the longitudinal coordinate for the electromagnetic field, while keeping the notation of $x$ for the coordinate of the electron. Applying the boundary conditions (thus setting $x^\prime=x$) one gets $E_i=E_r=\frac{1}{2}\cdot\sqrt{1+\omega_p^2}\cdot E_t$ and $\phi_r=2\phi_t=2(\alpha-x)$, where $\alpha\simeq\arctan\omega_p$ with $\omega_p=\sqrt{n_e}$ the plasma frequency. One can use the transmitted vector potential at $x=x^\prime$ to obtain the \emph{driving} vector potential:

\begin{equation}
a_y=-\frac{2E_i}{\sqrt{1+\omega_p^2}}\sin(t-x +\alpha)\cdot e^{-\omega_p(x^\prime-x )}\mathrm{.}
\label{Adrequation}
\end{equation}

It is important to notice that the actual vector potential driving the electron is approximately $\omega_p/2$ times lower than the incident one. As a consequence relativistic effects to the electron motion and corresponding corrections to the skin depth become important only when the amplitude of the incoming light $a_0$ exceeds $\omega_p/2$ \cite{Wilks93}.

\begin{figure}
\includegraphics[width=0.45\textwidth]{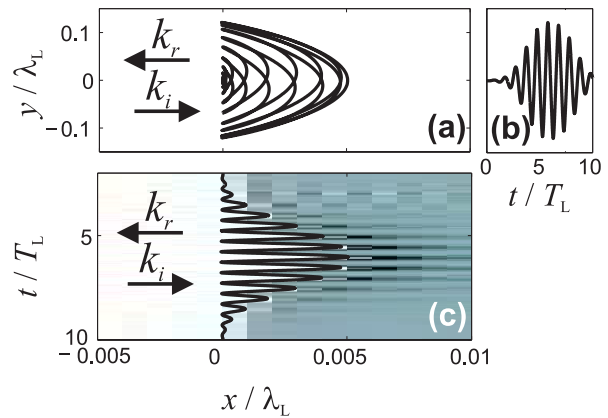}
\caption{Electron motion obtained using the capacitor model for laser pulse with $a_0=10$ with 4 cycles FWHM-duration and $n_e=400$. Electron is initially located at $x_e=y_e=0$. Subfigure (a) shows the electron trajectory, subfigure (b) demonstrates the behaviour of the transverse coordinate $y_e$ in time, on subfigure (c) the dashed line represents the longitudinal coordinate $x_e$ of the electron (vertical axis) versus time (horizontal axis) obtained from the model, the color coded image displays the spatio-temporal picture of the electron density obtained from 1D-PIC simulations with same laser and plasma parameters.\label{modelmotion}}
\end{figure}

Results of the model calculations are presented in Fig.~\ref{modelmotion}. The trajectory of an electron  interacting with a laser pulse having Gaussian envelope of 4-cycles Full Width Half Maximum (FWHM) duration $\tau_{FWHM}$ and amplitude $a_0=10$ (corresponding to an intensity of $1.37\cdot 10^{20}$ W/cm$^2$ for a laser wavelength of $\lambda_L=1$ $\mu$m) is shown in Fig.~\ref{modelmotion}a. Plasma density is $n_e=400$. Fig.~\ref{modelmotion}b shows the transverse coordinate $y$ as a function of time $t$. In Fig.~\ref{modelmotion}c the solid line shows the longitudinal coordinate $x$ (horizontal axis) as a function of time $t$ (vertical axis).

Figure ~\ref{modelmotion}c allows us to understand the origin of the harmonic generation process. One can see that during the interaction the model-electron (the step-like reflecting surface) oscillates in longitudinal direction with twice the laser frequency. Each time the surface moves towards the laser it produces a flash with attosecond duration \cite{Tsa06NJP}. This simple and intuitive picture is called the Oscillating Mirror model and was proposed by Bulanov {\it et al.} \cite{Bulanov94} and further developed by Lichters {\it et al.} \cite{Lichters96}. In this paper we want to pay attention to the transverse motion of the electron, which extends to a considerable fraction of the laser wavelength (see Fig.~\ref{modelmotion}b) and therefore might be responsible for the surface smoothing. Indeed if the transverse motion of the electron exceeds the characteristic size of the modulations on a rough surface, then the roughness is likely to disappear.

In assessing the role of the transverse motion as a possible smoothing mechanism a  simple expression for its amplitude is needed. Neglecting longitudinal motion one can get from Eq. \ref{Adrequation} an estimate for the amplitude of the transverse motion $y_{max}$:
\begin{equation}
y_{max}\approx\frac{2\cdot a_0}{\sqrt{\omega_p^2+4a_0^2}}
\label{ymaxequation}
\end{equation}

\begin{figure}
\includegraphics[width=0.48\textwidth]{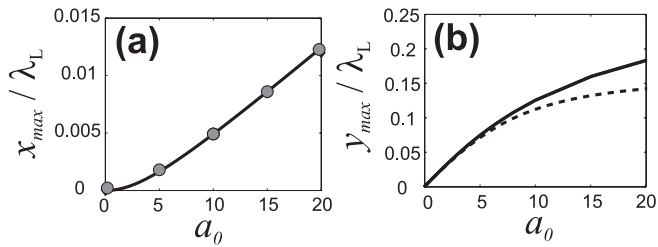}
\caption{Dependance of the amplitudes of longitudinal $x_{max}$(a) and transverse $y_{max}$ (b) motion on laser amplitude $a_0$. In figure (a) the circles represent the results of 1D PIC simulations and the solid line shows the results of the numerical integration of the capacitor model. In figure (b) the solid line depicts the numerical integration of the model equations and the dashed line is obtained from equation (\ref{ymaxequation}). \label{modelascan}}
\end{figure}

The dependence  of the transverse electron motion  amplitude $y_{max}$ on the laser pulse amplitude $a_0$ for plasma density $n_e=400$ is shown on Fig.~\ref{modelascan}b. The solid line shows the results obtained by numerically solving the model equations and the dashed line represents equation (\ref{ymaxequation}). The simple estimate (\ref{ymaxequation}) works fairly good for the parameter range studied, and its simplicity makes it convenient for the following estimates. More accurate results can be obtained by numerically integrating the model equations.

Having estimated the amplitude of the transverse coordinate $y_{max}$ one can establish an \emph{ad-hoc} criterion for surface smoothing to occur based on the ratio of this amplitude to the characteristic roughness size $h$. For instances where the transverse motion is on the order of the characteristic roughness size within the interaction area, considerable smoothing can be expected. One can define a dimensionless parameter $\xi$ separating the case when smoothing takes place from the case when the roughness survives during the interaction:
\begin{equation}
\xi=\frac{2a_0}{\sqrt{\omega_p^2+4a_0^2}\cdot h_y}\cdot e^{-\omega_ph_x}\mathrm{,}
\label{roughnessparameter}
\end{equation}
where $h_x$ and $h_y$ are the characteristic roughness size in longitudinal and transverse directions respectively. We make an assumption that the boundary conditions stay the same independent of surface structure and that the field exponentially decays inside the plasma. In the case when $\xi\gg1$ the roughness according to our criterion vanishes. We show further that even in the case when $\xi\sim 1$ substantial smoothing is observed.

\begin{figure}[htb]
\includegraphics[width=0.45\textwidth]{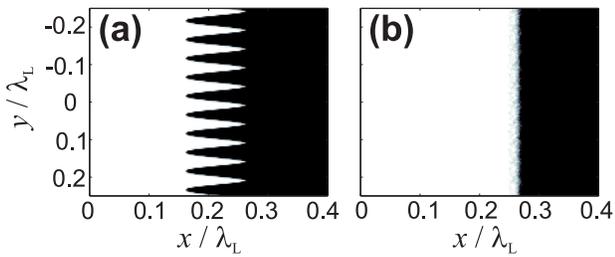}
\caption{(a)Initial density profile and (b) the smoothed density profile in the middle of the interaction process  for the surface with $\xi=0.6$.}
\label{denssmooth}
\end{figure}

In order to check the validity of the afore-described model and to demonstrate the surface smoothing we have conducted a series of 1D and 2D PIC simulations using the code PICWIG \cite{Rykovanov08} with clean and rough surfaces for different laser amplitudes $a_0$. The code allows the simulation of the interaction of the intense laser pulses with pre-ionized non-collisional plasma with the beam incident normally onto the target. The typical plasma density used in 1D simulations is $n_e=400$ and $n_e=30$ in the 2D case. A step-like vacuum-plasma interface is assumed, the ions are immobile. In the 2D case the surface is modulated sinusoidally in order to simulate the roughness (see left part of Fig.~\ref{denssmooth}). For convenience, the modulation period and amplitude are linked and the position of the vacuum-plasma interface is given by the law $x=h\cdot\sin(2\pi{y}/{h})$. The laser pulse amplitude was varied up to $a_0=20$ in the 1D case and is fixed to $a_0=10$ in the 2D scenario. Throughout the paper we use FWHM of the electric field as the definition of the laser pulse duration and use pulses with an electric field that has a Gaussian envelope function in both time and space.
\begin{equation}
E_y(t,x,y)=E_0\cdot\exp\left[-\frac{y^2}{2\rho^2} \right]\exp\left[-\frac{(t-x)^2}{2\tau_L^2} \right]\mathrm{,}
\label{pulsefunction}
\end{equation}
where $\rho$ and $\tau_L$ are the width of the focus and duration of the laser pulse respectively. The FWHM duration is related to $\tau_L$ by $\tau_{FWHM}=\tau_L\sqrt{8\ln2}$. In the 1D case the size of the simulation box is 7$\lambda_L$, the time step is $T_L/1000$ with $T_L$ the period of the driving laser and each plasma cell is initially occupied by 1000 macro-electrons. In the 2D case the size of the simulation box is 3.5$\lambda_L$ in laser propagation direction and 40$\lambda_L$ in polarization direction. The time step is $T_L/300$ and the laser propagation direction spatial step is $\lambda_L/300$. Each cell is initially occupied by 50 macro-electrons.

\begin{figure*}[htb]
\includegraphics[width=0.98\textwidth]{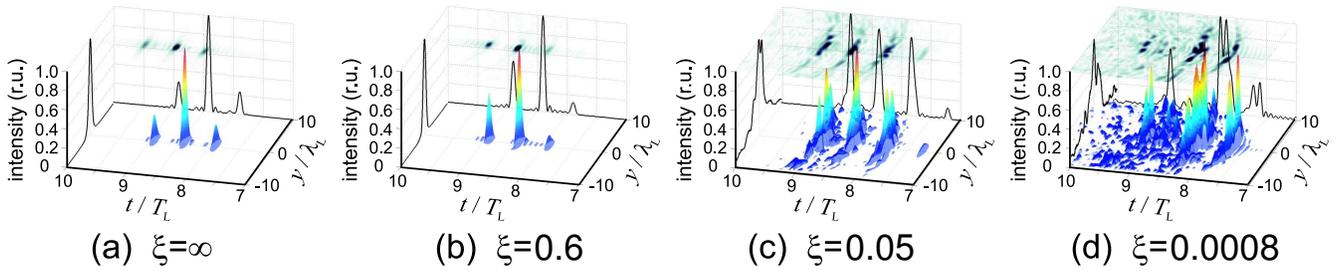}
\caption{Farfield distribution of the reflected harmonics beam $200\lambda_L$ away from the target for a) smooth surface, b) surface with modulation size $h=0.05\lambda_L$, c) surface with modulation size $h=0.1\lambda_L$, d) surface with modulation size $h=0.2\lambda_L$.}
\label{farfield}
\end{figure*}

The results of 1D simulations are presented on Fig.~\ref{modelmotion}c and Fig.~\ref{modelascan}a. The color-coded image on Fig.~\ref{modelmotion}c presents the spatio-temporal evolution of the electron density obtained from simulations with same laser and plasma parameters as in the model (solid line). One can see that the model is in agreement with the PIC simulations. Fig.~\ref{modelascan}a shows results of the model calculations of longitudinal electron amplitude (solid line) for different laser amplitudes $a_0$ compared to simulations (circles). The fact that simulation results lie on the curve obtained from the model and as longitudinal motion is directly correlated to transverse motion allows us to claim that the model works well and gives correct results for both longitudinal (Fig.~\ref{modelascan}a) and transverse (Fig.~\ref{modelascan}b) coordinates. The latter are hard to obtain from 1D PIC simulation as the particles leave the interaction region and are very intricate to trace.

In the 2D case we investigate the spatial beaming of harmonics as a possible indication of smoothing. We analyze the propagation of the harmonics emission away from the target using Kirchhoff diffraction theory \cite{Born99} following the approach used in earlier investigations \cite{Geissler07, Hoerlein09}. The harmonic beam (from 15th to 25th harmonic, central wavelength $0.05\lambda_L$) 200 $\lambda_L$ away from the target is shown on Fig.~\ref{farfield}. On all four sub-figures the color surface presents the distribution of normalized intensity of the filtered harmonics as a function of both time $t$ and transverse coordinate $y$ (the ceiling panel shows the same data as a color-coded image). The upper-right plane shows the projection of the beam to the time axis thus the time structure of the harmonics beam exhibiting a train of several attosecond pulses. On the upper-left plane the intensity distribution of the harmonics beam as a function of transverse coordinate $y$ is shown (black solid line). Results presented on Fig.~\ref{farfield}~a,b,c,d are obtained for a surface with modulation size $h=0$ (smooth surface, $\xi\rightarrow\infty$), $h=0.05\lambda_L$ ($\xi\approx 0.6$), $h=0.1\lambda_L$ ($\xi \approx 0.05$) and $h=0.2\lambda_L$ ($\xi\approx 0.0008$) respectively. There are several important points to mention.

First, for the simulation parameters considered the distance of 200$\lambda_L$ corresponds to the position of the harmonics focus due to surface denting as discussed in the paper by H\"{o}rlein \emph{et al.} \cite{Hoerlein09}. This can be illustrated from the Fig.~\ref{farfield}a by the fact that the transverse width of the reflected harmonics beam (see graph in the upper-left plane) is much less than the initial laser width with $\rho=5 \lambda_L$.

Secondly for dimensionless smoothing parameter $\xi$ on the order of unity the spatial and temporal structure of the harmonic beam is not influenced by the surface roughness. Figure \ref{farfield} shows the harmonic orders from the 15th to the 25th, which should undergo diffuse reflection by each of the rough surfaces simulated. Contrary to the Rayleigh criterion \cite{Ishimaru78}, but in agreement with experimental observation \cite{Dro08Div}, almost no change in the harmonic beam structure is observed for $\xi\approx 0.6$ (Fig.~\ref{farfield}b) in good agreement with our \emph{ad-hoc} smoothing criterion.
Surfaces with $\xi\ll1$ (Fig.~\ref{farfield}c,d) lead to the speckle-like diffraction picture with more energy going to the wings of the beam. The fact that the beam is still tolerably collimated hints that even though $\xi\ll1$, the characteristic surface roughness was significantly diminished during the interaction. The analysis of the spatial structure of harmonics generated on the corrugated surfaces exhibits collimated beam structure and serves as an indirect proof of the surface smoothing.

 Direct proof of the surface smoothing can be found on Fig.~\ref{denssmooth} where initial density distribution (as function of longitudinal and transverse coordinates) and the density distribution near the moment when the pulse maximum reaches the surface are shown (left and right sub-figures respectively). The results here are presented for the surface with $\xi\approx 0.6$. The evolution of the electron density in time can be traced in the animation made from simulation data (see Supplementary material), showing that the transverse motion of the electrons leads to rapid (in contrast to the hydrodynamically slow smoothing due to ion motion) smoothing of the corrugation.

In conclusion we have shown for the first time that coherent electron dynamics is
the dominant effect shaping the laser plasma interaction surface. This is a paradigm
shift from the way that surface dynamics have been viewed to date - as purely
hydrodynamic in nature. Due to their much smaller mass, electrons can modify the
surface morphology on the time-scale of even the shortest, few-cycle laser pulses
and hence must be taken into account when considering intense laser-plasma
interactions. This effect has important consequences in the field of ultrafast
pulses  and their application (e.g. harmonic generation, ion acceleration via
radiation pressure) and implies that surface imperfections on a scale smaller than
the laser wavelength can be neglected.

\begin{acknowledgments}
This work was funded in part by the International Max-Planck Research School-APS, the MAP excellence cluster,  and by the Association EURATOM - MPI f\"ur Plasmaphysik.
\end{acknowledgments}


\end{document}